\PassOptionsToPackage{unicode}{hyperref}
\PassOptionsToPackage{hyphens}{url}
\documentclass[
]{article}
\usepackage{amsmath,amssymb}
\usepackage{iftex}
\ifPDFTeX
  \usepackage[T1]{fontenc}
  \usepackage[utf8]{inputenc}
  \usepackage{textcomp} 
\else 
  \usepackage{unicode-math} 
  \defaultfontfeatures{Scale=MatchLowercase}
  \defaultfontfeatures[\rmfamily]{Ligatures=TeX,Scale=1}
\fi
\usepackage{lmodern}
\ifPDFTeX\else
\fi
\IfFileExists{upquote.sty}{\usepackage{upquote}}{}
\IfFileExists{microtype.sty}{
  \usepackage[]{microtype}
  \UseMicrotypeSet[protrusion]{basicmath} 
}{}
\makeatletter
\@ifundefined{KOMAClassName}{
  \IfFileExists{parskip.sty}{%
    \usepackage{parskip}
  }{
    \setlength{\parindent}{0pt}
    \setlength{\parskip}{6pt plus 2pt minus 1pt}}
}{
  \KOMAoptions{parskip=half}}
\makeatother
\usepackage{xcolor}
\usepackage[margin=1in]{geometry}
\usepackage{graphicx}
\makeatletter
\def\maxwidth{\ifdim\Gin@nat@width>\linewidth\linewidth\else\Gin@nat@width\fi}
\def\maxheight{\ifdim\Gin@nat@height>\textheight\textheight\else\Gin@nat@height\fi}
\makeatother
\setkeys{Gin}{width=\maxwidth,height=\maxheight,keepaspectratio}
\makeatletter
\def\fps@figure{htbp}
\makeatother
\usepackage{listings}
\usepackage[numbers,sort&compress]{natbib}
\ifLuaTeX
  \usepackage{selnolig}  
\fi
\IfFileExists{bookmark.sty}{\usepackage{bookmark}}{\usepackage{hyperref}}
\IfFileExists{xurl.sty}{\usepackage{xurl}}{} 
\hypersetup{
  pdftitle={Non-parametric inference on calibration of predicted risks},
  pdfauthor={Mohsen Sadatsafavi*; John Petkau},
  pdfkeywords={Risk prediction; Inference; Hypothesis testing;
Calibration},
  hidelinks,
  pdfcreator={LaTeX via pandoc}}

\title{Non-parametric inference on calibration of predicted risks}
\author{Mohsen Sadatsafavi* \and John Petkau}
\date{May 2024}

\begin{document}
\maketitle
\begin{abstract}
Moderate calibration, the expected event probability among observations
with predicted probability z being equal to z, is a desired property of
risk prediction models. Current graphical and numerical techniques for
evaluating moderate calibration of risk prediction models are mostly
based on smoothing or grouping the data. As well, there is no widely
accepted inferential method for the null hypothesis that a model is
moderately calibrated. In this work, we discuss recently-developed, and
propose novel, methods for the assessment of moderate calibration for
binary responses. The methods are based on the limiting distributions of
functions of standardized partial sums of prediction errors converging
to the corresponding laws of Brownian motion. The novel method relies on
well-known properties of the Brownian bridge which enables joint
inference on mean and moderate calibration, leading to a unified
`bridge' test for detecting miscalibration. Simulation studies indicate
that the bridge test is more powerful, often substantially, than the
alternative test. As a case study we consider a prediction model for
short-term mortality after a heart attack, where we provide suggestions
on graphical presentation and the interpretation of results. Moderate
calibration can be assessed without requiring arbitrary grouping of data
or using methods that require tuning of parameters.
\end{abstract}

\let\thefootnote\relax\footnotetext{From Faculty of Medicine and Faculty of Pharmaceutical Sciences (Mohsen Sadatsafavi) and Department of Statistics (John Petkau), The University of British Columbia, Vancouver, BC, Canada}
\let\thefootnote\relax\footnotetext{*Correspondence to Mohsen Sadatsafavi, Room 4110, Faculty of Pharmaceutical Sciences, 2405 Wesbrook Mall, Vancouver, BC, V6T1Z3, Canada; email: msafavi@mail.ubc.ca}

\hypertarget{introduction}{%
\section{Introduction}\label{introduction}}

Calibration for a risk prediction model refers to the ability of the
model to generate predicted probabilities that are close to their true
counterparts. Calibration is a critical element of a risk model's
performance. It has been shown that the clinical utility of a risk
prediction model is more sensitive to its calibration than to its
discrimination \cite{VanCalster2015MDMCPMCalibDCA}. Poorly calibrated
predictions also provide incorrect information to patients and hinder
informed shared decision-making. Despite this, the critical role of risk
model calibration remains underappreciated, so much so that it is called
the Achilles' heel of predictive analytics
\cite{VanCalster2019AchillesHeel}.

For the prediction of binary outcomes, Van Calster et
al.~\cite{VanCalster2016calibrationHierarchy} proposed a hierarchy of
calibration definitions. According to this hierarchy, mean calibration
(aka calibration-in-the-large) refers to the closeness of the average of
predicted and observed risks, and is often the very first step of
evaluating a model \cite{Steyerberg2014ABCD}. Weak calibration refers to
the intercept and slope of the `calibration line', obtained by fitting a
logit model associating the response with logit-transformed predicted
risks, being equal to, respectively, zero and one. The corresponding
likelihood ratio (LR) test enables inference on weak calibration.
Moderate calibration refers to the average risk among all individuals
with a given predicted risk being equal to that predicted risk. Finally,
strong calibration requires that the predicted and observed risks are
equal within each subgroup defined by all distinct covariate patterns.
The authors argue that strong calibration is neither achievable nor
strictly desirable, and that moderate calibration should be the ultimate
metric of merit for risk prediction models.

Moderate calibration is often visually assessed via a flexible
calibration plot (henceforth referred to as calibration plot for
brevity) \cite{VanCalster2016calibrationHierarchy}, which is the plot of
the conditional mean of observed risk (y-axis) as a function of
predicted risk (x-axis). When the response is binary, in the common
situation that predicted risks are continuous with no or few ties,
calculating this conditional mean requires grouping the data into bins
or applying local smoothing methods. Moderate calibration is then
assessed by the closeness of this conditional mean to the line of
perfect calibration (identity line). Scalar indices have been proposed
for summarizing this discrepancy. Examples include Emax, E50, Estimated
Calibration Index \cite{VanHoorde2015ECI} and Integrated Calibration
Index (ICI) \cite{Harrell2015RegressionBook, Austin2019ICI}. As such
metrics require estimating the conditional mean function, their
calculation similarly requires a tuning parameter (e.g., loess
bandwidth).

In addition to visual assessment, answering the question
``\textit{is this model moderately calibrated in this population?}'' can
be approached as a formal inference problem. The historical approach
towards inference on moderate calibration for binary outcomes has been
the Hosmer-Lemeshow (HL) test. This test is based on grouping predicted
risks into quantiles (conventionally deciles) and comparing predicted
and observed event frequencies, with the resulting statistic
approximated as chi-square \cite{Hosmer1980HLTest}. The HL test has been
criticized due to its sensitivity to the arbitrary specification of the
number of groups and lack of information it provides on the direction of
miscalibration \cite{Hosmer1997GoFComparison}. An alternative test for
moderate calibration has been proposed that does not require tuning
parameters, but it is simulation-based and thus is subject to Monte
Carlo error \cite{Sadatsafavi2022mROC}.

Hosmer et al.\cite{Hosmer1997GoFComparison} reviewed several alternative
goodness-of-fit (GOF) tests for binary responses. However, none of the
alternative methods examined seems to be a test for moderate
calibration. They defined the purpose of GOF testing for logistic
regression models to be three-fold: to examine whether the logit
transformation is the correct link function, whether the linear
predictor is of the correct form (i.e., there is no need for the
inclusion of additional covariates, transformations, or interactions),
and whether the variance of the response variable conditional on
covariates is that of Bernoulli distribution
\cite{Hosmer1997GoFComparison}. Similarly, Allison
\cite{Allison2014GoFComparison} considered the objective of GOF testing
to be to examine whether a more complicated model is required. These are
fundamentally different pursuits than examining moderate calibration.

Given this, classical methods do not provide an opportunity for an
objective and unequivocal assessment of moderate calibration. In this
work, we discuss recently proposed, and propose novel, methods for
inference on moderate calibration and associated metrics for quantifying
the degree of miscalibration, that do not involve regularization. Our
novel method is motivated by standard practice in predictive analytics
which involves evaluating mean calibration before, or in tandem with,
moderate calibration \cite{Steyerberg2014ABCD}. As mean calibration is a
necessary condition for moderate calibration, proper inference on
moderate calibration should incorporate the results of mean calibration
evaluation. As we demonstrate, such a two-pronged approach also
increases statistical power.

The rest of this manuscript is structured as follows: after describing
the context, we briefly review recently proposed graphical and
statistical methods for moderate calibration based on partial sums of
prediction errors. Motivated by these developments, we construct a
stochastic process that converges asymptotically to standard Brownian
motion, enabling us to take advantage of known properties of Brownian
motion to propose joint inference on mean and moderate calibration. We
conduct simulation studies comparing the performance of these tests. We
showcase these methods in a case study in prediction of mortality after
a heart attack, where we propose graphical illustration of results and
interpretation of findings. We conclude by discussing how the proposed
test can be added to the toolbox for examining model calibration, and
suggesting future research in this area.

\hypertarget{methods}{%
\section{Methods}\label{methods}}

\hypertarget{notation-and-context}{%
\subsection{Notation and context}\label{notation-and-context}}

A risk prediction model can be conceptualized as a function \(\pi(X)\)
mapping a covariate pattern X into a predicted risk \(\pi\). We have
access to a representative sample of \(n\) independent individuals
randomly selected from a target population. Let
\(\pi=(\pi_1,\pi_2,…,\pi_n )\) be the set of predicted risks, ordered
from smallest to largest. We assume there are no ties (e.g., at least
one of the predictors is continuous) and no extreme predictions
\((\forall i<j: 0<\pi_i<\pi_j<1)\). Let \(Y=(Y_1,Y_2,…,Y_n )\) be the
corresponding observed binary outcomes, taking values in \{0,1\}. Our
task is to test the null hypothesis (\(H_0\)) that the model is
moderately calibrated; that is, \(\mathbb{E}(Y|\pi=z)=z\) for any
predicted risk \(z\). We consider \(\pi\)s fixed at their calculated
values \cite{Delgado1993RegCurvesEquality}.

\hypertarget{a-stochastic-process-approach-towards-the-assessment-of-moderate-calibration}{%
\subsection{A stochastic process approach towards the assessment of
moderate
calibration}\label{a-stochastic-process-approach-towards-the-assessment-of-moderate-calibration}}

Our proposed approach is informed by fundamental theories in
non-parametric model evaluation based on partial sums of residuals
\cite{Delgado1993RegCurvesEquality, Diebolt1995NonparametricGoF, Stute1997NonparametricGoF},
which have recently been tuned to the specific case of evaluating the
calibration of models for binary outcomes
\cite{Tygert2020Plots, ArrietaIbarra2022BM}.

The calibration plot is the plot of an estimate of the conditional mean
\(f(z)=\mathbb{E}(Y|\pi=z)\) versus \(z\). Evaluating moderate
calibration is equivalent to evaluating whether \(f(z)=z\). In the
common situation that the \(\pi\)s are generated from a continuous
distribution, estimating \(f(z)\) requires regularization (e.g., binning
or smoothing). However, one can also examine moderate calibration on the
partial sum domain. Consider the partial sums of predictions,
\(\sum_{i=1}^{n} \pi_iI(\pi_i \leq z)\), and the corresponding partial
sums of responses, \(\sum_{i=1}^{n} Y_iI(\pi_i \leq z)\) , where \(I()\)
is the indicator function. It follows immediately from the definition of
moderate calibration that
\(\mathbb{E} \big[ \sum_{i=1}^{n} Y_iI(\pi_i \leq z) \big] = \sum_{i=1}^{n} \pi_iI(\pi_i \leq z)\)
under \(H_0\) for all \(z\)s. Evaluating this equality for any fixed
\(z\) no longer requires regularization as, after scaling by \(n\), the
expectation can be consistently estimated from the data.

This is the core concept in the recently proposed methods for the
assessment of calibration by Tygert \cite{Tygert2020Plots} and
Arrieta-Ibarra et al.~\cite{ArrietaIbarra2022BM}. These papers consider
the behavior of the (scaled) partial sums of prediction errors:

\begin{equation}
C_i = \frac{1}{n}\sum_{j=1}^i(Y_j-\pi_j) \text{, for } i=1,2,3,...,n.
\end{equation}

\noindent Tygert \cite{Tygert2020Plots} proposed plotting \(C_i\) on the
y-axis as a function of \(i/n\) (with \(\pi_i\) on a secondary x-axis).
Arrieta-Ibarra et al.~\cite{ArrietaIbarra2022BM} took advantage of the
functional Central Limit Theorem (CLT) applied to such partial sums to
develop statistical tests for calibration. They focused on the maximum
absolute value and maximum range. In what follows, we focus on the
maximum absolute value given the enhanced interpretability of the
resulting statistic for clinical prediction models (as will be explained
later). The quantity of interest is

\begin{equation}
\label{Cstar}
C^* = \max_{1 \leq i \leq n} |C_i|,
\end{equation}

\noindent which was referred to as ECCE-MAD (maximum absolute deviation
of the empirical cumulative calibration error) by Arrieta-Ibarra et
al.~\cite{ArrietaIbarra2022BM}. They argued that under \(H_0\), a
properly scaled version of this quantity, i.e.,

\begin{equation}
S^* = \frac{nC^*}{\sqrt {\sum_{i=1}^n \pi_i(1-\pi_i)}},
\label{Sstar}
\end{equation} converges in distribution to the supremum of the absolute
value of standard Brownian motion \(W(t)\) over the unit interval {[}0,
1{]}, thus acting as a test statistic whose cumulative distribution
function (CDF) under \(H_0\) can be computed via the rapidly convergent
series \cite{Borodin1996BMHandbook}

\begin{equation}
F(a) = P(\sup_{0 \leq t  \leq 1}|W(t)| < a)=4/\pi\sum_{k=0}^{\infty} \frac{(-1)^k}{2k+1}e^{-\frac{(2k+1)^2\pi^2}{8a^2}}.
\label{BMCDF}
\end{equation}

\noindent From this, one can generate an asymptotic p-value for moderate
calibration as \(p=1-F(S^*)\). We will hereafter refer to this test as
the Brownian motion (BM) test.

In addition to enabling inference, this partial sum process provides
interpretable metrics of miscalibration. Indeed \(C_n\) is the mean
calibration error, a consistent estimator of \(\mathbb{E}(Y-\pi)\). As
well, \(C^*\) converges to
\(\sup_{0 \leq t \leq 1} |\int_0^t e(z)dP(z)|\) , where
\(e(z)=\mathbb{E}(Y|\pi=z)-z\) is the local calibration error at \(z\)
and \(P(z)\) is the limit of the empirical CDF of the predicted risks.
For comparison, ICI is an estimator of \(\int_0^1 |e(z)|dP(z)\), while
Emax is an estimator of \(\sup_{0 \leq z \leq 1}|e(z)|\). In this vein,
as a distance quantity, \(C^*\) combines elements from both ICI and
Emax, with the advantage that its estimation does not require any tuning
parameter.

\hypertarget{a-stochastic-process-that-converges-to-standard-brownian-motion}{%
\subsection{A stochastic process that converges to standard Brownian
motion}\label{a-stochastic-process-that-converges-to-standard-brownian-motion}}

Arrieta-Ibarra et al.'s development was based on certain functionals of
the partial sums having the limiting laws as those of Brownian motion.
We strengthen the connection to Brownian motion by explicitly
constructing a stochastic process that asymptotically converges at all
points to Brownian motion. Consider the random walk defined by the set
of \(n+1\) \{time, location\} points:

\begin{equation}
\begin{cases}
  t_i = \frac{1}{T}\sum_{j=1}^i \pi_j(1-\pi_j)\\
  
  S_i = \frac{1}{\sqrt T}\sum_{j=1}^i (Y_j-\pi_j) = \frac{n}{\sqrt T}C_i
\end{cases}
\end{equation}

\noindent with \(i=0,1,\ldots,n, t_0=0, S_0=0,\) and
\(T=\sum_{i=1}^{n}{\pi_i(1-\pi_i)}\). Note that \(t_n=1\).

Under \(H_0\), the \(i\)th time step in this random walk is proportional
to \(var(Y_i)=\pi_i(1-\pi_i)\), and the vertical jumps are standardized
such that \(var(S_n)=1\). The motivation for this setup is that under
\(H_0\), \(\mathbb{E}(S_i)=0\), \(var(S_i)=t_i,\) and
\(cov(S_i,S_j)=\min(t_i,t_j)\). These are the main features of standard
Brownian motion \(W(t)\) on the time axis \(t\). \(\{S_i\}\) is a
martingale, as \(\mathbb{E}(S_i|S_0,S_1,\ldots, S_{i-1})=S_{i-1}\); as
well, the centered Bernoulli random variables \(Y_j-\pi_j\) satisfy the
Lindeberg condition \cite{Lindeberg1922LindebergCondition}. These two
conditions allow the application of Brown's Martingale CLT, which
establishes that the continuous-time stochastic process created by
linearly interpolating the points \((t_i,S_i)\) converges weakly to
\(W(t)\) on the {[}0,1{]} time interval \cite{Brown1971MartingaleCLT}.

We note that the maximum absolute value of this random walk,
\(S^*=\max_{1 \leq i \leq n}(S_i)\), is identical to Arietta-Ibarra's
test statistic in (\ref{Sstar}), thus immediately obtaining the same
limiting distribution as in (\ref{BMCDF}). In addition, the terminal
value, \(S_n=nC_n/\sqrt T\), is directly related to mean calibration
and, under \(H_0\), \(S_n\) converges in law to
\(W(1)\sim Normal(0,1)\).

\hypertarget{a-novel-test-for-moderate-calibration-based-on-the-properties-of-brownian-bridge}{%
\subsection{A novel test for moderate calibration based on the
properties of Brownian
bridge}\label{a-novel-test-for-moderate-calibration-based-on-the-properties-of-brownian-bridge}}

The weak convergence of the above stochastic process to \(W(t)\) enables
taking advantage of the known properties of the latter in devising
potentially more powerful inference methods. For example, given that
mean calibration is a necessary condition for moderate calibration, one
can consider \(P(S_n,S^*)\), the joint distribution of the terminal
value and maximum absolute value of the stochastic process. This can be
decomposed to \(P(S_n)P(S^*|S_n)\). As mentioned above, under \(H_0\),
\(S_n\) converges in law to \(W(1)\) which has a standard normal
distribution; that is, \(S_n\) is asymptotically a Z-score for mean
calibration. On the other hand, under \(H_0\), \(P(S^*|S_n)\) converges
to the distribution of the supremum of the absolute value of standard
Brownian motion conditional on a given terminal value. The corresponding
CDF can be expressed as the following rapidly convergent series
\cite{Borodin1996BMHandbook}:

\begin{equation}
\label{conditionalCDF} 
P(\sup_{0 \leq t \leq 1}|W(t)|<a | W(1)=b)=\sum_{k=-\infty}^{\infty} (-1)^k e^{2abk-2a^2k^2}.
\end{equation}

These results can be the basis of a two-part test for moderate
calibration, with a marginal test based on \(S_n\) and a conditional
test based on \(S^*\) given \(S_n\). However, use of the latter means
assessment of the significance of \(S^*\) depends on the observed
\(S_n\),complicating its interpretation. A more interpretable, and
potentially more powerful, approach would be one based on two
independent statistics.

To progress, we propose a new summary statistic for moderate calibration
whose asymptotic distribution under \(H_0\) is independent of that of
\(S_n\). In doing so we take advantage of the properties of Brownian
bridge. A Brownian bridge, \(B(t)\), is a Brownian motion conditional on
\(W(1)=0\), and can be constructed from a given realization of Brownian
motion \(W(t)\) via the transformation \(B(t)=W(t)-tW(1)\)
\cite{Chow2009BrownianBridge}. An attractive property of the Brownian
bridge is that the path of \(B(t)\) constructed by transforming \(W(t)\)
as above is independent of \(W(1)\): for any \(t\), \(B(t)\) and
\(W(1)\) are bivariate-normal random variables and
\(cov(B(t),W(1))=cov(W(t),W(1))-t\,var(W(1))=0\). Thus, examining if the
`bridged' random walk \(S_i-t_iS_n\) behaves like a Brownian bridge will
provide an asymptotically independent opportunity from \(S_n\) for
evaluating moderate calibration. In Appendix A, we show that the two
components of such a test asymptotically guarantee moderate calibration.

Correspondingly, we decompose \(H_0\) into two components of examining
the terminal position of the random walk (\(H_{0A}\)) and the deviation
of the bridged random walk from its expected path under \(H_0\)
(\(H_{0B}\)). For \(H_{0A}\), an asymptotic two-tailed p-value can be
generated via a Z test: \begin{equation}
p_A=2\Phi(-|S_n|),      
\end{equation}

\noindent where \(\Phi\) is the standard normal CDF. In practice a
t-test might be used, with variance estimated from the data. However, as
long as there is more than trivial number of observations, which is
usually satisfied in practical applications, the two tests will generate
nearly identical p-values.

A natural choice for examining \(H_{0B}\) is the maximum absolute value
of the bridged random-walk: \begin{equation}
B^*=\max_{1 \leq i \leq n}(|S_i-t_i S_n|), 
\end{equation}

\noindent which, under \(H_0\), converges in law to the supremum of the
absolute value of the Brownian bridge, with the following rapidly
converging CDF \cite{Marsaglia2003Kolmogorov}:

\begin{equation}
G(a)=P(\sup_{0 \leq t \leq 1} |B(t)| <a)=\sum_{k=-\infty}^{\infty}{(-1)^k e^{-2a^2k^2}}=\frac{\sqrt{2\pi}}{a}\sum_{k=1}^\infty{e^{-\frac{(2k-1)^2 \pi^2}{8a^2}}}.
\end{equation}

\noindent This is the CDF of the Kolmogorov distribution, the asymptotic
null distribution of the one-sample Kolmogorov-Smirnov test statistic.
The wide implementation of this CDF in statistical analysis software is
a further reason to favor this test over the conditional test based on
(\ref{conditionalCDF}). From this CDF, an asymptotic one-tailed p-value
can be calculated as

\begin{equation}
p_B=1-G(B^*).
\end{equation}

Given the independence of \(W(1)\) and \(B(t)\), the two asymptotic
p-values for \(H_{0A}\) and \(H_{0B}\) are independent. Consequently,
one can use Fisher's method of combining independent p-values to obtain
a unified asymptotic p-value for moderate calibration
\cite{Littell1971FisherMethodOptimality}: \begin{equation}
p=1-CDF_{\chi^2_{(4)}}(-2\,log(p_A)-2\,log(p_B)).
\end{equation}

\noindent We refer to this test as the (Brownian) bridge (BB) test.
Appendix B provides an exemplary R implementation of this test.

\hypertarget{simulation-studies}{%
\section{Simulation studies}\label{simulation-studies}}

We conducted a series of simulation studies to evaluate the
finite-sample behavior of the Brownian motion and bridge tests under the
null hypothesis, and their comparative power in detecting various forms
of model miscalibration. All numerical results in this work are produced
in R \cite{RCoreTeam2019R}. Our implementation of \(F()\) is provided in
the \textit{cumulcalib} package (function
pMAD\_BM)\cite{Sadatsafavi2024cumulcalibPackage}. For the Kolmogorov
distribution (\(G()\)), we used the implementation in the \textit{CPAT}
package \cite{Miller2018RCPATPAckage}, but one can trick any
implementation of the asymptotic Kolmogorov-Smirnov test to obtain the
desired results (an example is provided in Appendix B).

\begin{figure}
\centering\includegraphics [width=0.75\textwidth]{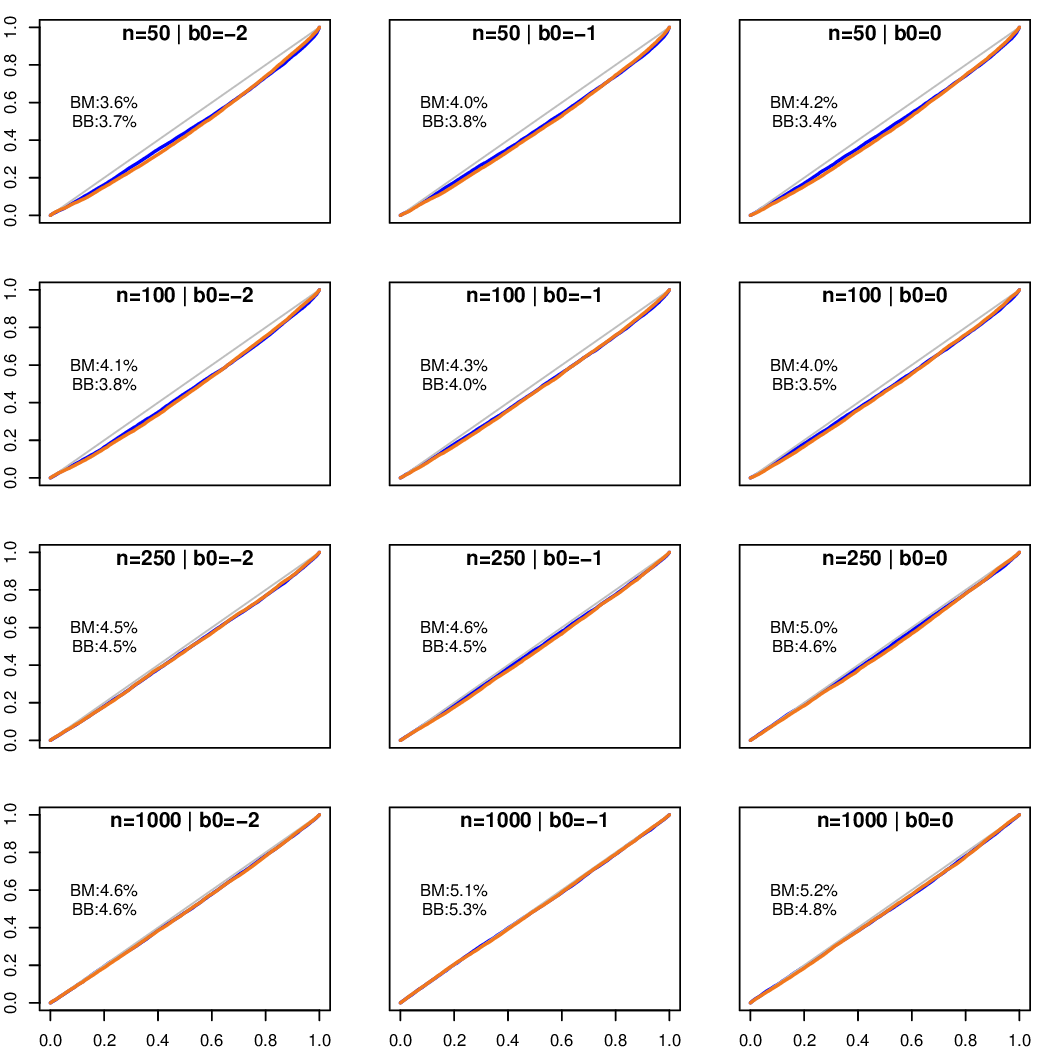}
\caption{Empirical cumulative distribution functions of p-values under null hypothesis. Blue: Brownian motion (BM) test; orange: Brownian bridge (BB) test; gray: identity line. Numbers on each panel are the proportion of times p-values exceeded 0.05.}
\label{fig:nullSimResults}
\end{figure}

The first set of simulations evaluated the finite-sample properties of
both tests under \(H_0\). This was performed to assess at what effective
sample size the null behavior of these tests is substantially different
from their asymptotic behavior. These simulations were performed by
generating predicted risks as \(\mbox{logit}(\pi)=\beta_0+X\), with
\(X\sim Normal(0,1)\), and responses \(Y\sim Bernoulli(\pi)\). We varied
\(\beta_0\) in \(\{-2,-1,0\}\) (resulting in average event probabilities
of, respectively, 0.155, 0.303, and 0.500) and sample size in
\(\{50,100,250,1000\}\) in a fully factorial design. Each scenario was
run 10,000 times. Empirical CDF plots of p-values are provided in Figure
\ref{fig:nullSimResults}. As p-values should follow the standard uniform
distribution under \(H_0\), empirical CDFs should be close to the
identity line. The proportion of simulated p-values\textless0.05 is also
reported on each panel. Generally, both tests behaved similarly. In
small samples, the p-values seem to be slightly biased upwards, so the
corresponding tests are conservative. The proportion of
p-values\textless0.05 was within 10\% of the desired proportion (i.e.,
0.045-0.055) only in the bottom half of the panels (\(n\geq250\)).
Taking this as a subjective criterion, the total variance
\(\sum_{i=1}^n \pi_i (1-\pi_i)\) (as a measure proportional to the
effective sample size) should be \(\geq30\) for the asymptotic behavior
to be acceptable. This corresponds to 40, 51, and 72 events for
\(\beta_0\) of -2, -1, and 0, respectively. However, we leave deeper
investigation of the small-sample properties of the tests to future
studies and also note that for the Kolmogorov distribution small-sample
corrections are available
\cite{Marsaglia2003Kolmogorov, Vrbik2020Kolmogorov}, although the
appropriate form of such a correction is not obvious for the current
context. No such correction was used here to provide more comparable
results between the two tests. In general, if any doubt exists about the
adequacy of the sample size, one can carry out a simulation-based
version of these tests where the null distribution of \(S^*\) and
\(B^*\) (and \(S_n\)) is obtained by simulating many vectors of
responses in each \(Y_i\sim Bernoulli(\pi_i)\).

Next, we evaluated the power of the tests in detecting miscalibration.
Following recommendations on objectively deciding on the number of
simulations \cite{Morris2019SimulationForStats}, we obtained the results
through 2,500 Monte Carlo samples such that the maximum S.E. of the
estimated probability of rejecting \(H_0\) would be 0.01. In these
simulations, we modeled a single predictor \(X\sim Normal(0,1)\) and the
true risk as \(p=1/(1+\mbox{exp}(-X))\). We evaluated the performance of
the tests in simulated independent samples of \(n\) observations when
the predicted risks suffered from various degrees of miscalibration.

\begin{figure}
\centering\includegraphics [width=0.75\textwidth]{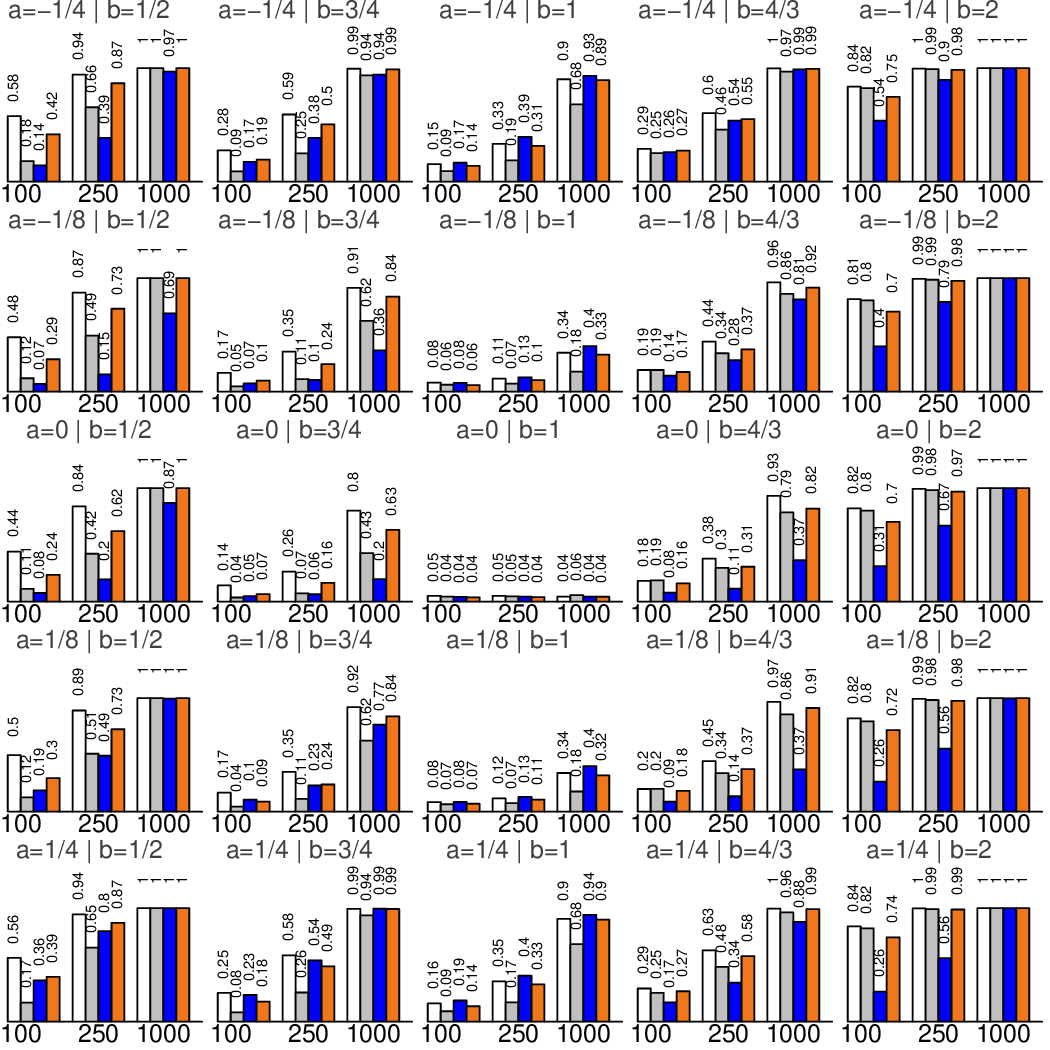}
\caption {Simulated power based on logit-linear miscalibration. White: Likelihood Ratio test, Gray: Hosmer-Lemeshow test (based on grouping by deciles), Blue: Brownian motion (BM) test, Orange: Brownian bridge (BB) test}
\label{fig:simResultsLinear}
\end{figure}

Two sets of simulations were performed. In the first set, we assumed the
risk model generated potentially miscalibrated predictions in the form
of \(\mbox{logit}(\pi)=a+b\,\mbox{logit}(p)=a+bX\). Given the linear
association on the logit scale between predicted and actual risks, weak
and moderate calibration are equivalent in these scenarios and,
therefore, the LR test (simultaneously testing whether \(a=0\) and
\(b=1\)), which is otherwise a test for weak calibration, has the
maximum theoretical power in detecting miscalibration. As such, this
simple setup provides an opportunity to judge the performance of the
tests against a gold standard. We simulated response values and
predicted risks in a fully factorial design with values
\(a=\{-1/4,-1/8,0,1/8,1/4\}\) and \(b=\{1/2,3/4,1,4/3,2\}\), and for
three sample sizes: \(n=\{100,250,1000\}\). Results are provided in
Figure \ref{fig:simResultsLinear}. All tests rejected the null
hypothesis around 5\% when the model was calibrated (the central panel).
As expected, the LR test had the highest power in this set of
simulations. In almost all scenarios, the bridge test had the highest
power among the other tests. One exception was when \(b=1\) (middle
column). In such instances, miscalibration is entirely in the mean
calibration component (\(a\)), and thus the bridge component of the
bridge test cannot be expected to contribute to the sensitivity of the
assessment.

\begin{figure}
\centering\includegraphics [width=0.75\textwidth]{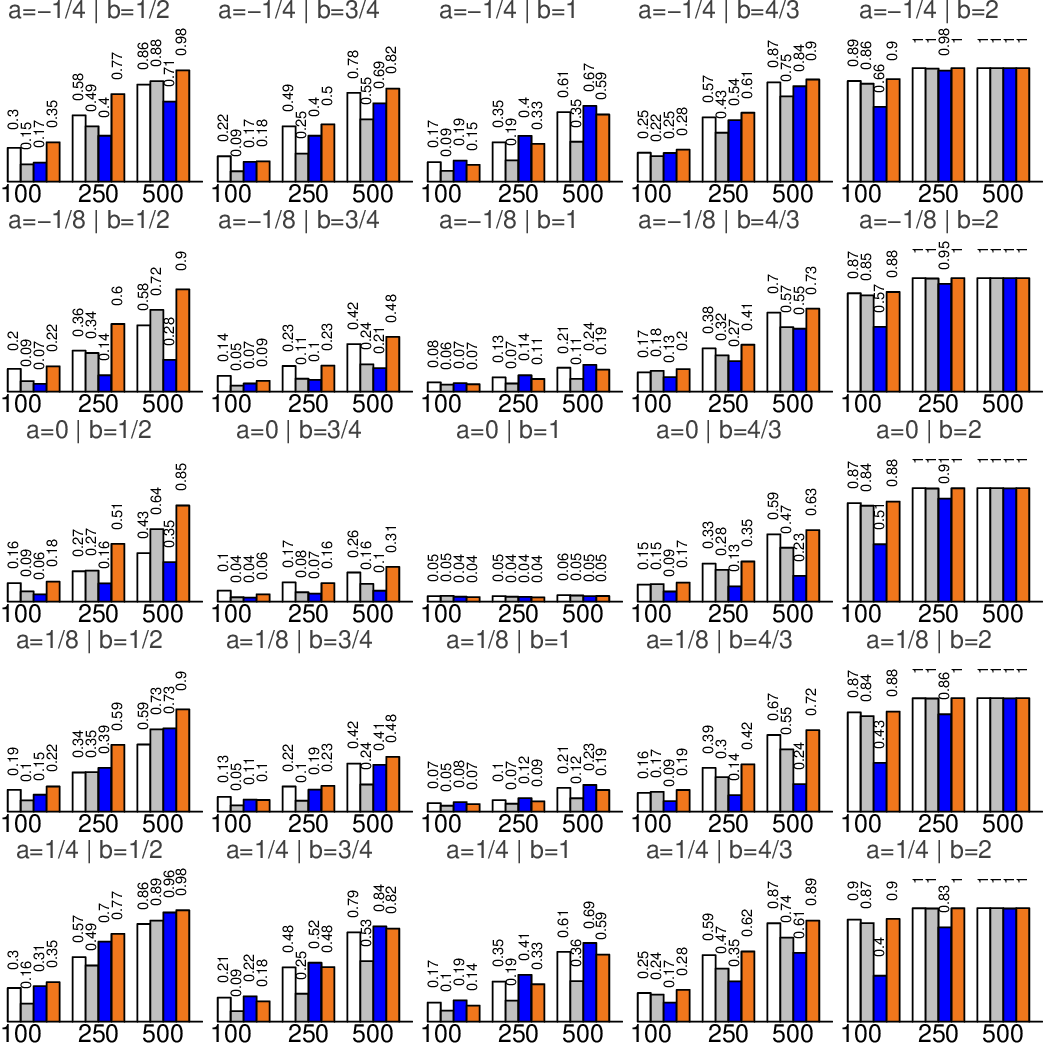}
\caption{Simulated power based on logit-power miscalibration. White: Likelihood Ratio test, Gray: Hosmer-Lemeshow test (based on grouping by deciles), Blue: Brownian motion (BM) test, Orange: Brownian bridge (BB) test}
\label{fig:simResultsPower}
\end{figure}

In the second set, the true risk model remained the same as above, and
we modeled non-linear miscalibrations as
\(\mbox{logit}(\pi)=a+b\,\mbox{sign}(X)|X|^{1/b}\). Here, \(a\) affects
mean calibration, while the term involving \(b\) is an odd function that
flexibly changes the calibration slope. Similar to the previous setting,
we simulated response values and predicted risks with values
\(a=\{-1/4,-1/8,0,1/8,1/4\}\), \(b=\{1/2,3/4,1,4/3,2\}\), and
\(n=\{100,250,1000\}\), in a fully factorial design. Figure
\ref{fig:simResultsPower} presents the results. Here, the LR test could
have low power due to non-linear miscalibrations (e.g., panels on the
left side of the figure). The bridge test was more powerful than the
Brownian motion test in almost all scenarios, again except those where
the shape of the calibration plot was unchanged (\(b=1\)) and only the
calibration intercept was affected (\(a\neq 1\)).

\hypertarget{case-study}{%
\section{Case study}\label{case-study}}

\begin{figure}
\centering
  \begin{tabular}{@{}c@{}}
    \includegraphics [width=0.5\linewidth,height=100pt]{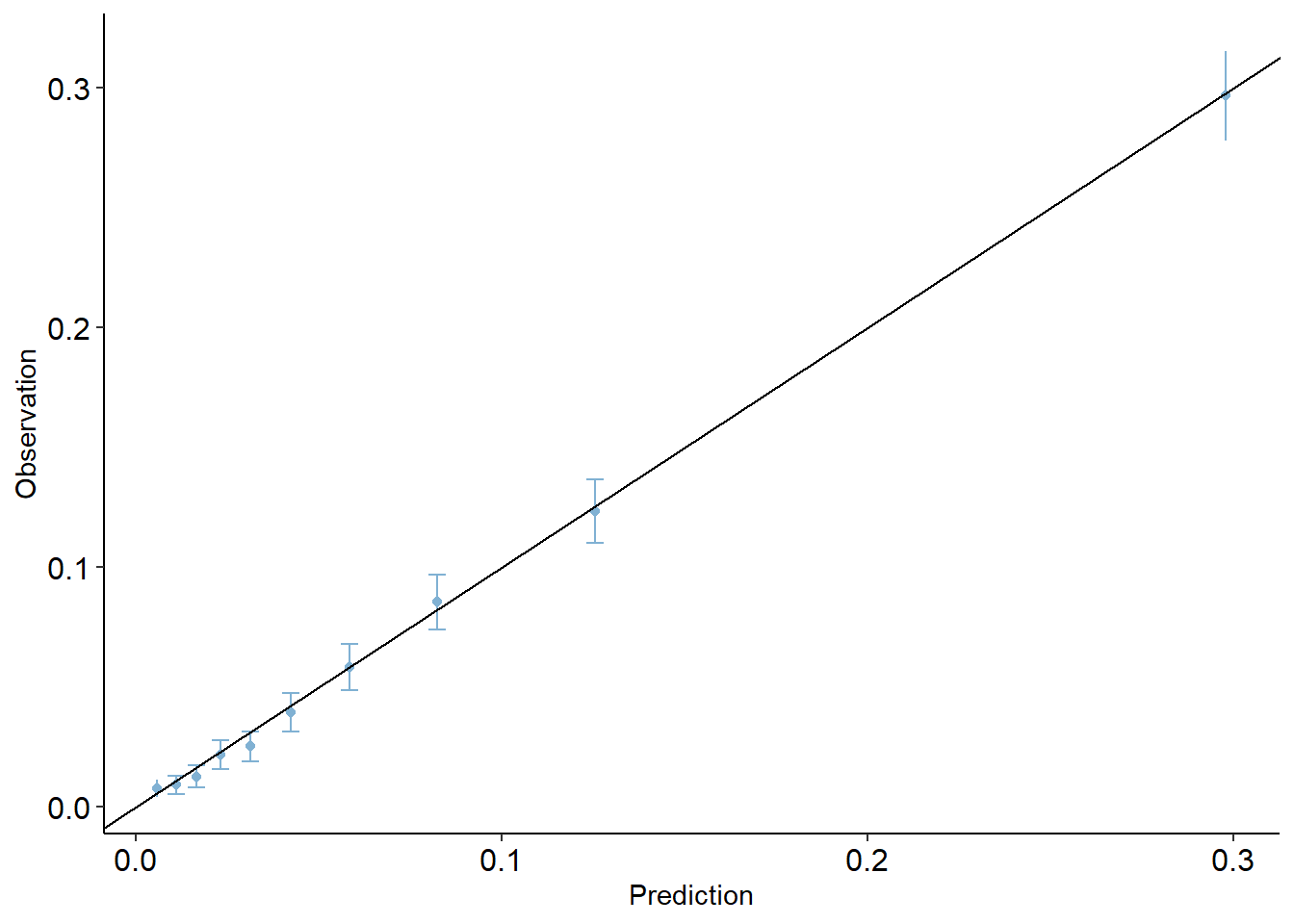} \\ [\abovecaptionskip]
    \small (a) Calibration plot (binned)
  \end{tabular}
  \begin{tabular}{@{}c@{}}
    \includegraphics [width=0.5\linewidth,height=100pt]{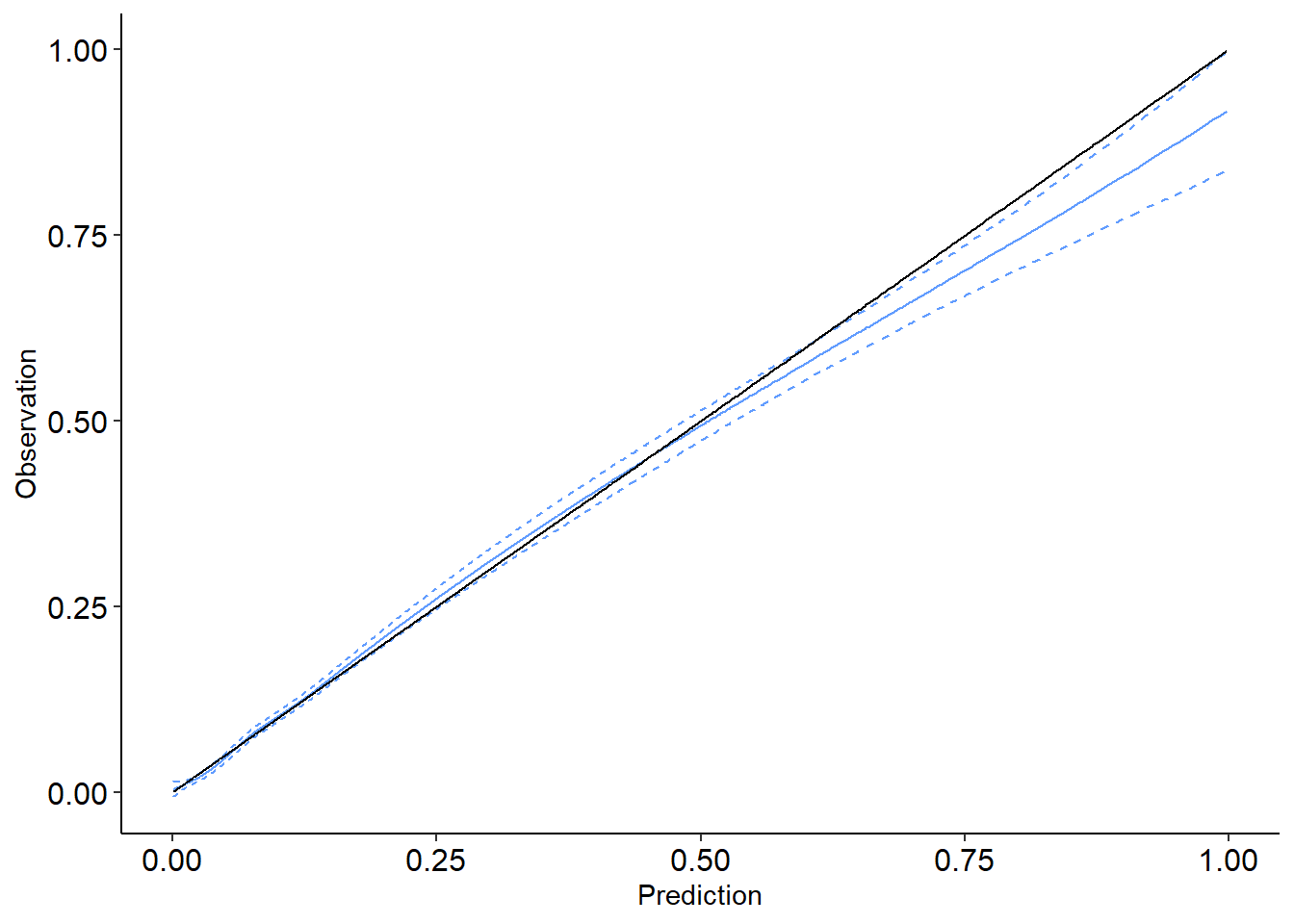} \\ [\abovecaptionskip]
    \small (b) Calibration plot (smoothed)
  \end{tabular}
  \begin{tabular}{@{}c@{}}
    \includegraphics [width=0.5\linewidth,height=140pt]{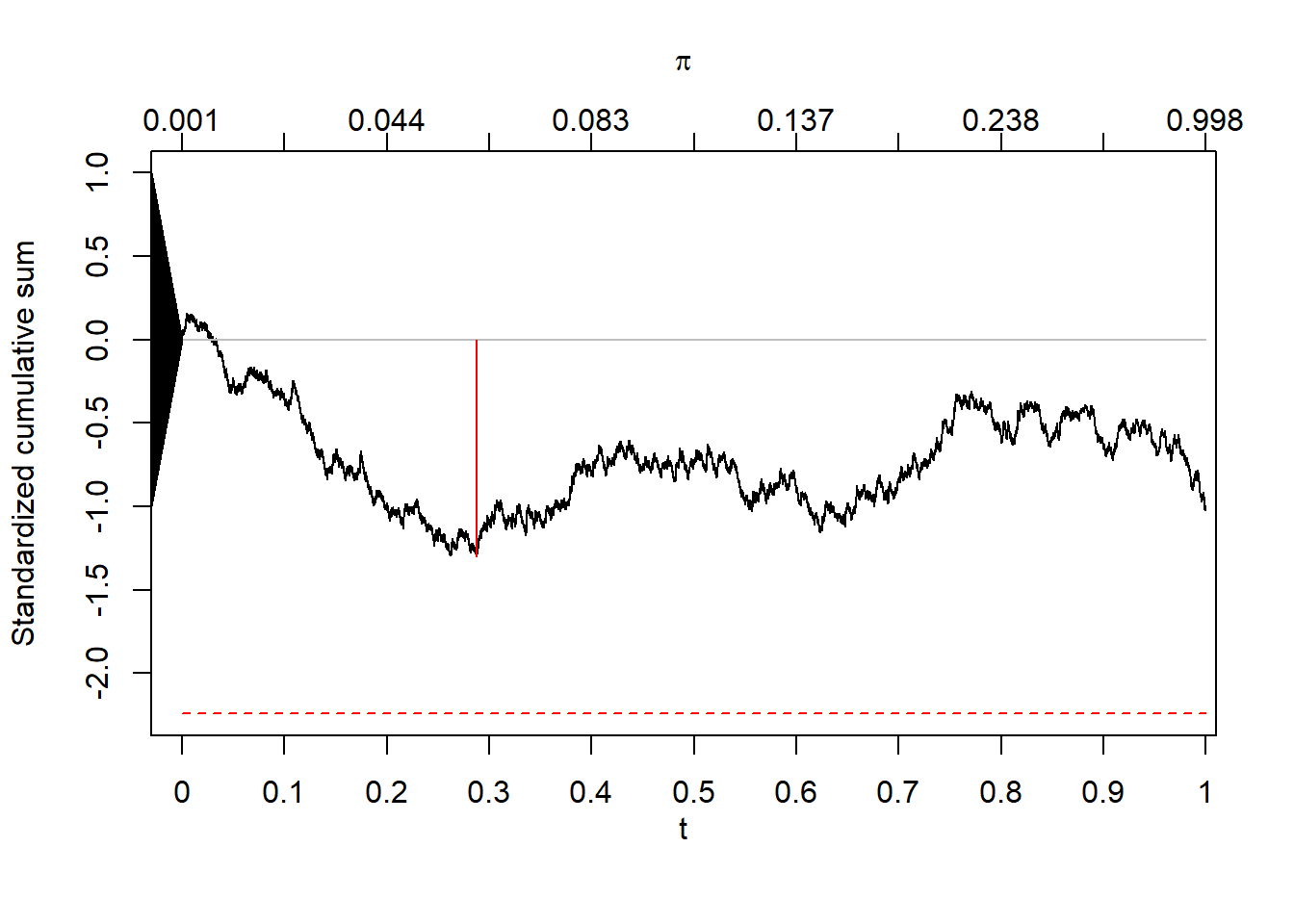} \\ [\abovecaptionskip]
    \small (c) Cumulative calibration plot (with Brownian motion test)
  \end{tabular}
  \begin{tabular}{@{}c@{}}
    \includegraphics [width=0.5\linewidth,height=140pt]{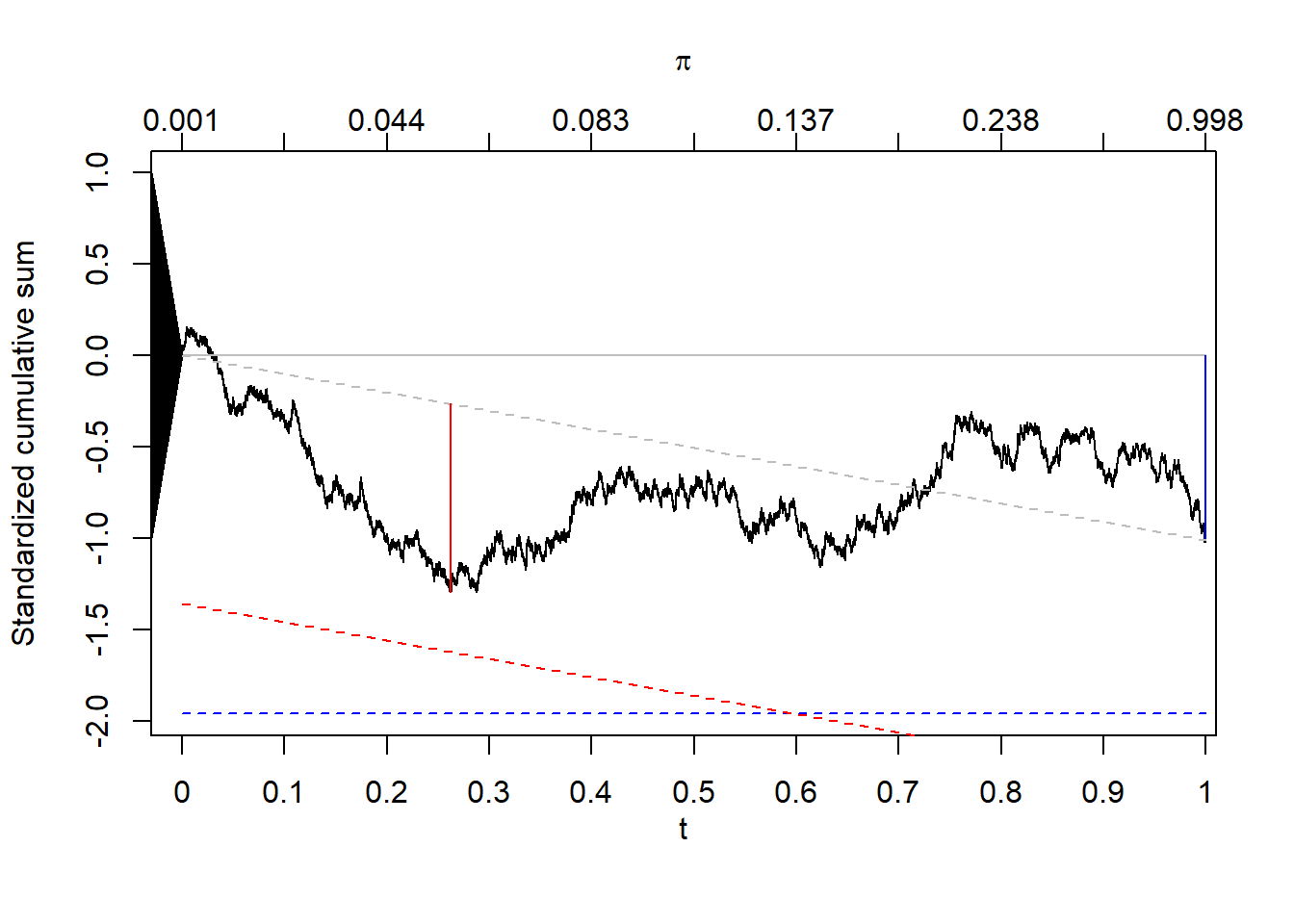} \\ [\abovecaptionskip]
    \small (d) Cumulative calibration plot (with Brownian bridge test)
  \end{tabular}
\caption {Calibration plots based on grouping by deciles (a) and based on loess smoothing (b) and the graph of the random walk $\{S_i\}$  (c and d), with the presentation of the Brownian motion (c) and bridge (d) tests for the full model. \newline \textit {Regression coefficients: {$\beta_0=-2.084,\beta_1= 0.078,\beta_2= 0.403,\beta_3= 0.577,\beta_4= 0.468,\beta_5= 0.767,\beta_6=-0.077,\beta_7= 0.018$}}}
\label{fig:caseStudyBigModel}
\end{figure}

\begin{figure}
\centering
  \begin{tabular}{@{}c@{}}
    \includegraphics [width=0.5\linewidth,height=100pt]{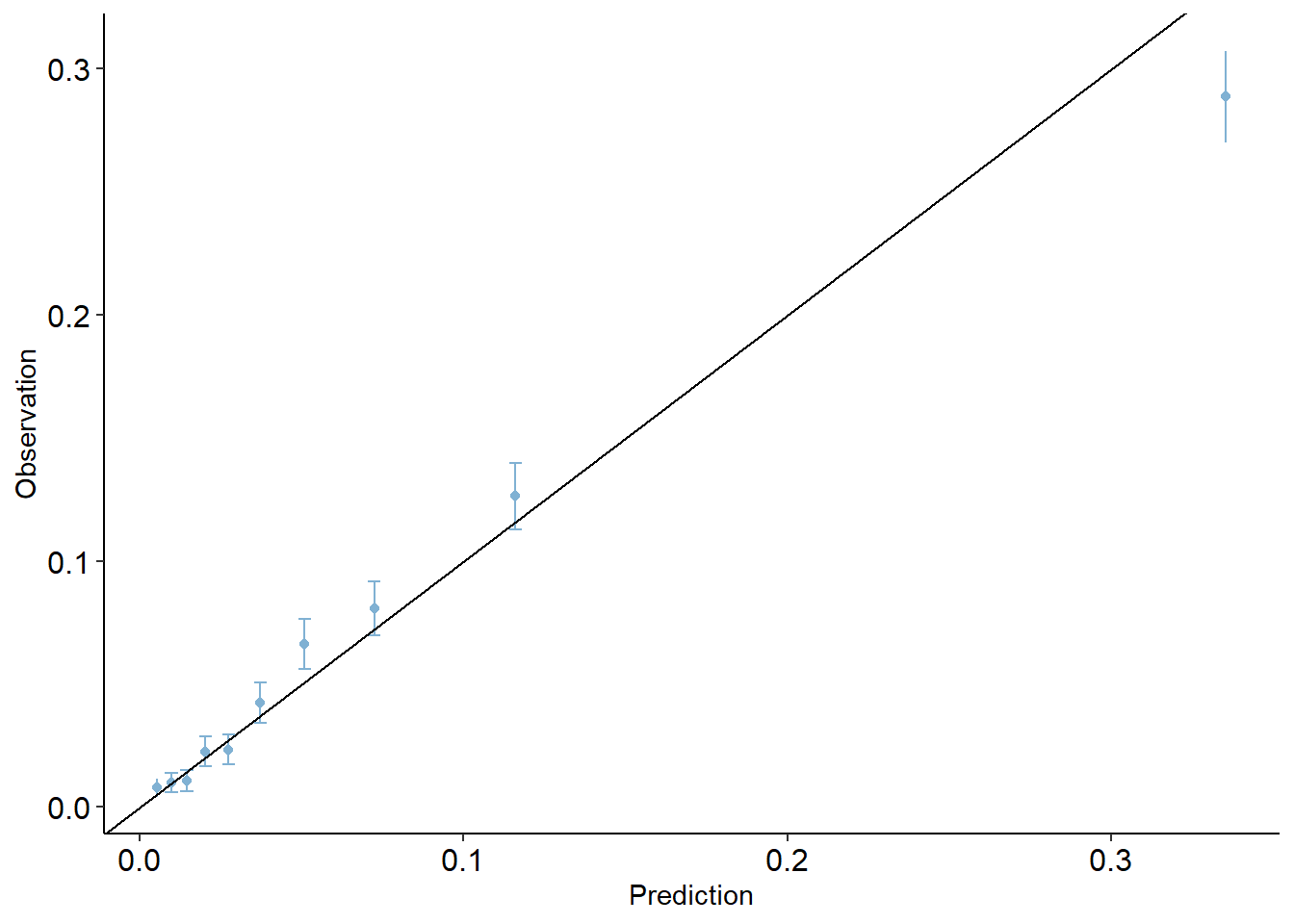} \\ [\abovecaptionskip]
    \small (a) Calibration plot (binned)
  \end{tabular}
  \begin{tabular}{@{}c@{}}
    \includegraphics [width=0.5\linewidth,height=100pt]{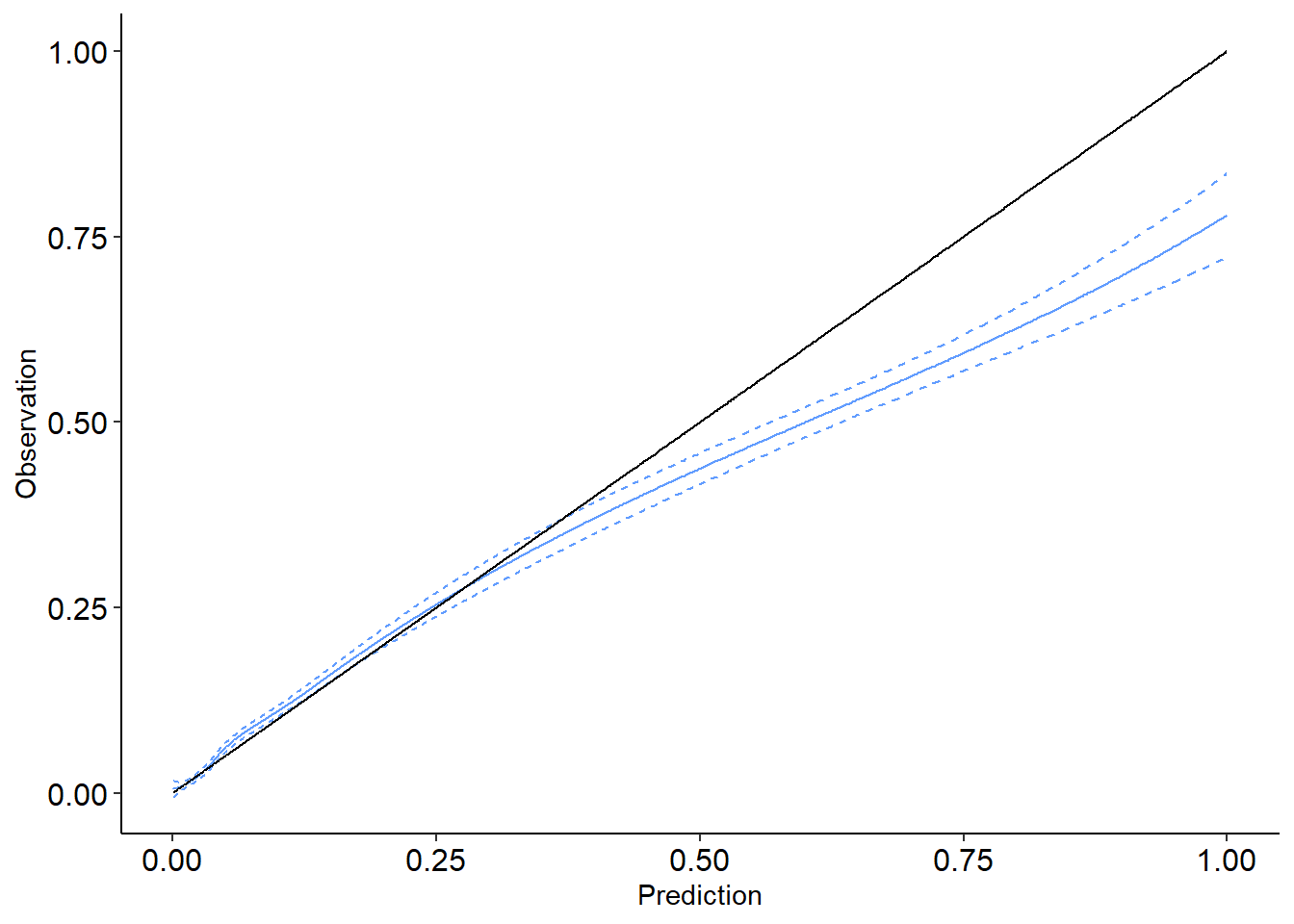} \\ [\abovecaptionskip]
    \small (b) Calibration plot (smoothed)
  \end{tabular}
  \vspace{\floatsep}
  \begin{tabular}{@{}c@{}}
    \includegraphics [width=0.5\linewidth,height=140pt]{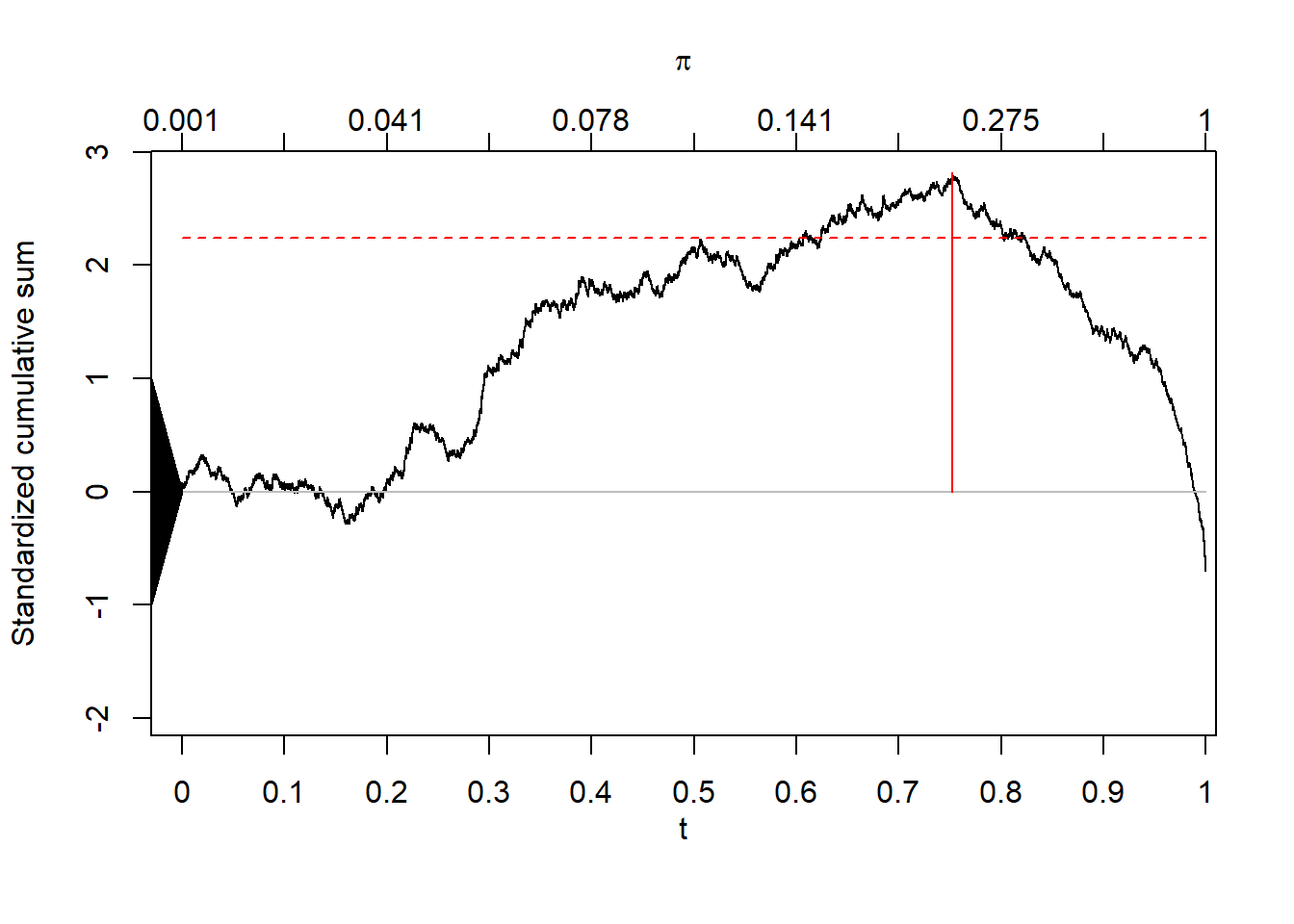} \\ [\abovecaptionskip]
    \small (c) Cumulative calibration plot (with Brownian motion test)
  \end{tabular}
  \vspace{\floatsep}
  \begin{tabular}{@{}c@{}}
    \includegraphics [width=0.5\linewidth,height=140pt]{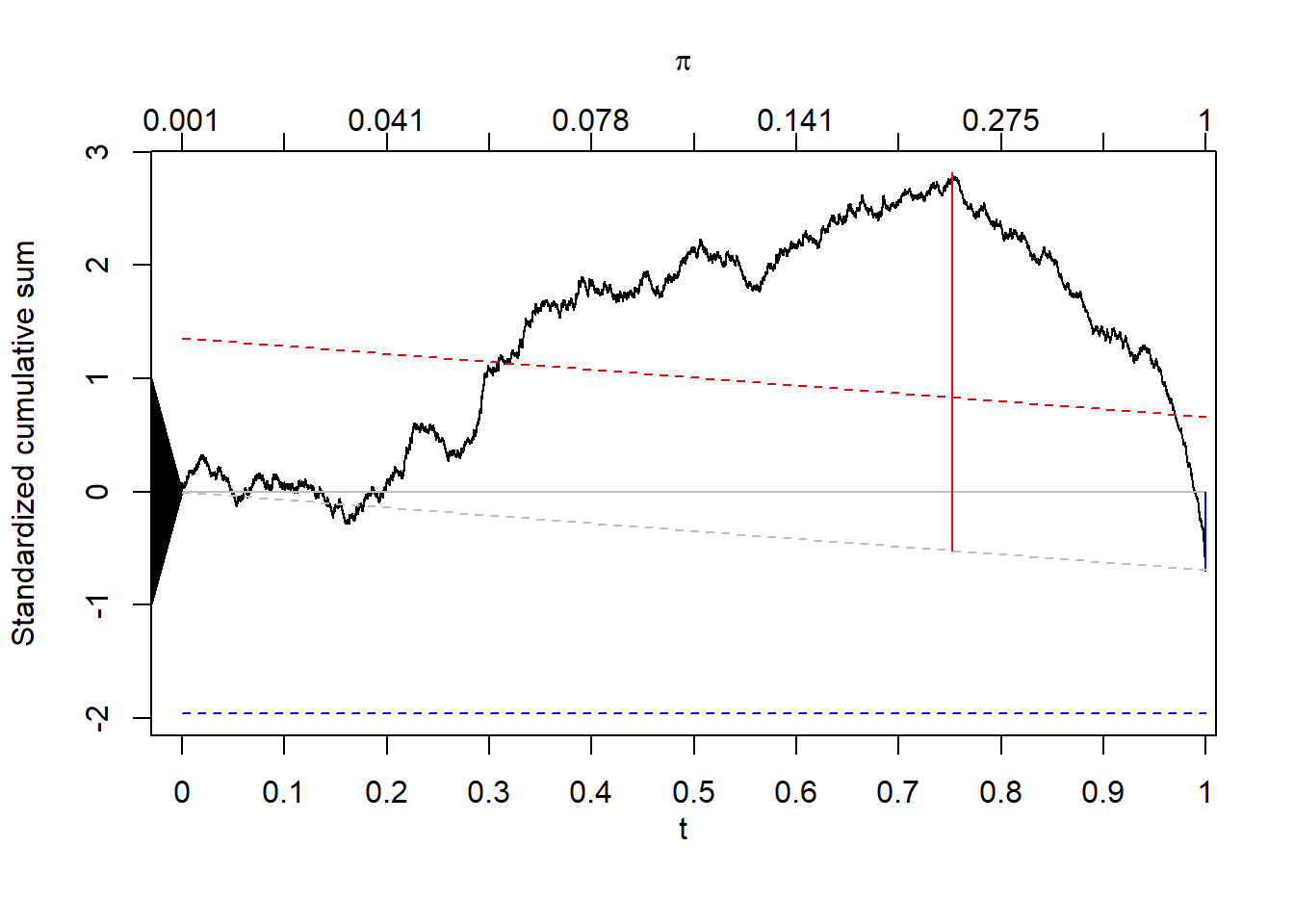} \\ [\abovecaptionskip]
    \small (d) Cumulative calibration plot (with Brownian bridge test)
  \end{tabular}
\caption {Calibration plots based on grouping by deciles (a) and based on loess smoothing (b) and the graph of the random walk $\{S_i\}$  (c and d), with the presentation of the Brownian motion (c) and bridge (d) tests for the small-sample model. \newline \textit {Regression coefficients: {$\beta_0= 0.365,\beta_1= 0.076,\beta_2= 0.460,\beta_3= 0.157,\beta_4= 0.349,\beta_5= 1.351,\beta_6=-0.103,\beta_7= 0.020$}}}
\label{fig:caseStudySmallModel}
\end{figure}

We used data from GUSTO-I, a clinical trial of multiple thrombolytic
strategies for acute myocardial infarction (AMI), with the primary
endpoint being 30-day mortality \cite{GUSTO1993RCT}. This dataset has
frequently been used to study methodological aspects of developing or
validating risk prediction models
\cite{Ennis1998GustoLearningMEthods, Steyerberg2005AMICPMs, Steyerberg2005AMICPMComparison}.
In line with previous studies, we used the non-US sample (\(n\)=17,796)
of GUSTO-I to fit a prediction model for 30-day post-AMI mortality, and
the US sample (\(n\)=23,034) for validating the prediction model
\cite{Sadatsafavi2022EVPICPMExVal}. We developed two models: one based
on the entire development sample, and one based on a subset of 500
observations from that sample (to create a potentially miscalibrated
model due to small development sample size). 30-day mortality was 7.2\%
in the full development sample, 8.2\% in the small development sample,
and 6.8\% in the validation sample. All the analyses were conducted in R
\cite{RCoreTeam2019R}. Ethics approval was not required because the
anonymized data are publicly available for research. Our candidate risk
prediction model is similar to those previously developed using this
dataset \cite{Sadatsafavi2022EVPICPMExVal}. The models were of the
following form (coefficients are provided in the footnote of Figures
\ref{fig:caseStudyBigModel} and \ref{fig:caseStudySmallModel})

\begin{center}
$\mbox{logit}(\pi)=\beta_0+\beta_1 [\text{age}]+\beta_2 [\text{AMI location other}]+\beta_3 [\text{AMI locaton anterior}]+\beta_4 [\text{previous AMI history}]+\beta_5 [\text{Killip score}]+\beta_6min [\text{blood pressure}],100)+\beta_7 [\text{heart rate}]$
\end{center}

\noindent (AMI location is dummy-coded, with posterior as the reference
category, previous AMI history is any previous AMIs, Killip score is a
measure of the severity of heart failure due to AMI). The c-statistic of
the full and small-sample models in the validation sample were,
respectively, 0.814 and 0.810.

Figure \ref{fig:caseStudyBigModel} -- panels (a) and (b) are calibration
plots of the full model, which seem to be compatible with good
calibration. Panels (c) and (d) present the graph of the \(\{S_i\}\)
process over time \(t\), with additional information pertaining to the
Brownian motion and the bridge tests, respectively. The corresponding
\(\pi\)s are provided as the secondary x-axis on top. To indicate the
size of the fluctuations, Tygert \cite{Tygert2020Plots} suggested adding
a triangle, centered at origin, with height equal to the standard
deviation of the terminal value of the random walk being plotted as an
aide for viewing the fluctuations of the process. We also find this
useful here, even though as our random-walk is standardized, the height
of the triangle will always be fixed at 1. In panel (c), the vertical
(red) line indicates the location and value of the test statistic
\(S^*\). In addition, if hypothesis testing is intended, the critical
value can be highlighted on the graph (the dashed red horizontal line,
for a 5\% significance level in this case). In our example, the line
indicates that the null hypothesis (that the model is moderately
calibrated) is not rejected at this level. \(C^*\) was \(0.0020\); the
test statistic (\(S^*\)) was \(1.2973\), corresponding to a p-value of
\(0.3889\). Panel (d) is our suggested presentation of the bridge test.
Here, the gray `bridge' line connecting the start and end of the random
walk is provided, and \(S_n\) and \(B^*\) are drawn as vertical
distances from this line (in, respectively, blue and red), along with
the above-mentioned triangle and dashed lines of significance (at 5\%
for each component - note the line for \(B^*\) is parallel to the bridge
line). The mean calibration, \(C_n\), was \(-0.0016\); the corresponding
Z-score (\(S_n\)), was \(-1.0091\) (p=\(0.3129\)), while the bridge test
statistic, \(B^*\), had a value of \(1.0284\) (p=\(0.2407\)). The bridge
test's unified p-value was \(0.2701\), indicating no substantial
miscalibration of this prediction model.

Figure \ref{fig:caseStudySmallModel} -- panells(a) and (b) are the
calibration plots of the small-sample model. The model seems to be
making optimistic predictions: underestimating the risk among low/medium
risk groups and overestimating it among high-risk groups. This is
clearer in the partial sum plot (panels (c) and (d)), which shows an
inverse U-shape that is typical of optimistic predictions. Again, the
Brownian motion and bridge tests are shown on the graph in panels (c)
and (d), respectively. \(C^*\) was \(0.0042\), corresponding to the test
statistic (\(S^*\)) of \(2.8151\) and a p-value of \(0.0098\). Maximum
cumulative prediction error occurred at \(\pi^*=0.232\) (\(t=0.752\)).
This is an estimation of the location (\(\pi\)) where the calibration
function \(\mathbb{E}(Y|\pi)\) crosses the identity line. Around this
point, predictions are calibrated. For individuals with predicted risk
above \(\pi^*\), the model overestimates the risk by an average of
\(0.086\) (note the downward slope of the partial sum on the right side
of the vertical red line in the middle panel). For others, it
underestimated the risk by an average of \(0.004\). Bridge test results
are shown in panel (d). Mean calibration was \(-0.0010\); the Z-score
for mean calibration was \(-0.6950\) (p=\(0.4871\)), while \(B^*\) was
\(3.3381\) (p\(<0.001\)). The unified p-value was \(<0.001\), clearly
indicating miscalibration of this prediction model.

\hypertarget{discussion}{%
\section{Discussion}\label{discussion}}

In this work, we built on recently proposed non-parametric methods for
the assessment of moderate calibration for models for binary outcomes.
We provided suggestions on visual and statistical assessment of model
calibration on the partial sum domain, proposed a novel non-parametric
test, and evaluated its performance against another recently proposed
test. A major advantage of these methods is no requirements for binning
or smoothing of observations. Our main innovation was a stochastic
process representation of the partial sum of prediction errors yielding
weak convergence to standard Brownian motion. This in turn enabled us to
propose a unified `bridge' test of moderate calibration based on two
asymptotically independent test statistics. These developments are
implemented as the R package
\textit{cumulcalib}\cite{Sadatsafavi2024cumulcalibPackage}.

Our simulations point towards the higher power of the (Brownian) bridge
test over the recently proposed Brownian motion test. This is not
surprising as the bridge test evaluates the behavior of the underlying
stochastic process in two independent ways, thus having more
opportunities for identifying miscalibration. However, the bridge test
also has other interesting properties. Asymptotically, \(B^*\) follows
the Kolmogorov distribution, which is ubiquitously implemented as part
of the Kolmogorov-Smirnov test. The joint inference on mean and moderate
calibration is also of particular relevance in clinical prediction
modeling, as oftentimes mean calibration is of independent relevance and
is the very first step in examining model calibration
\cite{Steyerberg2014ABCD}. As mean calibration is a necessary condition
for moderate calibration, the joint inference helps control the overall
error rate of inference on calibration. We used Fisher's method for
generating a unified p-value due to its optimality properties
\cite{Littell1971FisherMethodOptimality}. One can also use alternative
methods for retaining an overall familywise error, such as Bonferroni's
correction for each component of the test.

Our focus in this work was to propose inferential methods and evaluate
their performance. However, we do not suggest inattentive use of
p-values for making true / false conclusions at an arbitrary
significance level on whether a model has good calibration in a
population. Inferential methods quantify the compatibility of data with
the null hypothesis (moderate calibration in this case), and as such
complement the existing visual and numerical assessments of
miscalibration. As mentioned earlier, evaluation of calibration on the
partial sum domain also results in `distance' metrics (e.g., \(C^*\) or
ECCE-MAD as referred to by Arrieta-Ibarra et
al.~\cite{ArrietaIbarra2022BM}), similar to metrics such as Emax and ICI
for calibration plots, with convergence to a finite value that is
affected by the degree of miscalibration. \(C^*\) can thus be used in a
similar fashion to ICI and Emax, with the added benefit that its
estimation does not require any tuning parameters. However, there are
situations where one can reasonably expect that the model is calibrated,
and thus hypothesis-testing becomes particularity relevant. An example
is when the target population is known to be identical or very close to
the development population (e.g., another random sample from the same
population, or different study sites participating in a clinical trial
with a strict protocol). Another relevant context is after a model is
recalibrated in a new sample. Here, it is justified to expect that the
revised model can provide calibrated predictions, and \(H_0\) has a
plausible chance of being true. Strictly speaking, using the same data
to revise the model and evaluate its performance will result in
violation of the assumption underlying the proposed tests (that the
predicted risks are fixed quantities). Nevertheless, as long as the
degrees of freedom used in the model updating process is small (e.g.,
changing the intercept, rather than re-estimating coefficients
\cite{Steyerberg2004CPMUpdating}), the suggested inference procedures
would still provide useful insight. Indeed, this is a typical
approximation underlying many GOF tests that follow model fitting.
Interestingly, this is one instance where the \(B^*\) test can be used
in isolation: maximum likelihood estimation of generalized linear models
(e.g., logistic regression) guarantees that the average of predicted and
observed risks will be equal. In such instances, \(S_n=0\) by the
estimation process, and thus \(B^*\) can be used in isolation, to avoid
loss of statistical power for testing another component of the null
hypothesis. Rejection of the null hypothesis in this case would indicate
the need to modify the linear component (e.g., adding interaction terms)
or the link function. In this sense, the bridge test acts as a GOF test
for logistic regression.

There are several areas for future research. The weak convergence of the
random walk \(\{S_i\}\) to Brownian motion opens the door for possibly
more powerful tests. We identified two aspects of the stochastic process
that provided independent opportunities for examining moderate
calibration, but one can expand this logic towards comparing multiple
aspects. Examples of possible attributes include minimum, maximum,
range, or integrated square value of the scaled partial sum process. An
important extension of this method would be for other types of outcome,
including survival outcomes (as proposed, for example, for ICI
\cite{Austin2022ICICompetingRisk}). To what extent small sample
correction will improve the approximation of the Kolmogorov distribution
to that of the bridge test statistic \(B^*\) can be investigated (and
similar small-sample corrections can be devised for the Brownian motion
test). The finite sample performance of the distance metrics based on
partial sum of prediction errors versus metrics based on the calibration
plots should be investigated.

\hypertarget{conclusion}{%
\section{Conclusion}\label{conclusion}}

Non-parametric approaches based on partial sums of prediction errors
lead to visual, numerical, and inferential tools that do not require any
grouping or smoothing of data, and thus enable objective, unequivocal,
and reproducible assessments. Simulations show that the proposed bridge
test for moderate calibration has comparatively high power in detecting
miscalibration. We recommend such assessments to accompany conventional
tools (such as calibration plots and distance metrics) to provide a more
complete picture of the calibration of a risk prediction model in a
target population.

\newpage
\bibliographystyle{unsrtnat}
\bibliography{references.bib}

\newpage
\section*{Appendix A: Asymptotic sufficiency of the bridge test for moderate calibration}

Let \(e(z)=\mathbb{E}(Y|\pi=z)-z\) be the calibration error at predicted
risk \(z\). The model is moderately calibrated if \(\forall z: e(z)=0\).
Let \(P(.)\) be the cumulative distribution function of the infinite
population of predicted risks. The argument is for when \(P(.)\) is
strictly monotone in {[}0,1{]}.

In the general case where the null hypothesis is not known to hold, the
random walk constructed in the main text might have a non-zero expected
value (drift). Indeed, as \(n \rightarrow \infty\), at time \(t\),
\(\frac{1}{\sqrt n}\mathbb{E}(S(t)) \rightarrow \frac{1}{\sqrt \tau} \int_0^{\pi_t}e(x)dP(x)\),
where \(\tau=\int_0^1x(1-x)dP(x)\), and \(\pi_t\) is the unique solution
for \(\pi\) in \(\frac{1}{\tau}\int_0^\pi x(1-x)dP(x)=t\).
Correspondingly, the bridged random walk will have an expected value of:
\begin{equation}\nonumber
\frac{1}{\sqrt n} \mathbb{E}(S(t)-tS(1))\rightarrow \frac{1}{\sqrt \tau} \int_0^{\pi_t} e(x)dP(x)-\frac{t}{\sqrt \tau}\int_0^1 e(x)dP(x).
\end{equation}

The second component of the bridge test (\(H_{0B}\)) demands that
\(S(t)-tS(1)\) converges to a Brownian bridge, thus requiring it to have
zero expected value everywhere:

\begin{equation}\nonumber
\forall t: \int_0^{\pi_t} e(x)dP(x) = t \int_0^1 e(x)dP(x).
\end{equation}

Given the 1:1 relation between \(\pi_t\) and \(t\) and the monotonicity
of \(P(.)\), we can map this equality to the \(\pi\) domain:

\begin{equation}\nonumber
\forall\pi: \int_0^\pi e(x)dP(x)= \frac{\int_0^\pi x(1-x)dP(x)}{\int_0^1 x(1-x)dP(x)}\int_0^1 e(x)dP(x).
\end{equation}

\noindent Taking the first derivative with respect to \(\pi\) and
re-arranging terms, we have:

\begin{equation}\nonumber
\forall\pi:\frac{e(\pi)}{\pi(1-\pi)}=\frac{\int_0^1 e(x)dP(x)}{\int_0^1x(1-x)dP(x)}.
\end{equation}

\noindent The right-hand side is a constant so the solutions are of the
form \(e(\pi)=k\pi(1-\pi)\), and it is immediately obvious that any
\(k\) provides a valid solution. On the other hand, the hypothesis for
mean calibration, \(H_{0A}\), demands that \(\mathbb{E}e(\pi)=0\).
Therefore, the two null hypotheses together impose
\(\mathbb{E}{k\pi(1-\pi)}=0\), which can only hold for \(k=0\), which
implies moderate calibration.

\newpage
\section*{Appendix B: R code for the implementation of the bridge test}

We use birth weight data in the MASS package and an exemplary model for
predicting low birth weight. Below, as a generic implementation we use
the ks.test implementation of Kolmogorov CDF by creating a vector of n
(=100) observation whose empirical CDF has distance B*/sqrt(n) from the
CDF of the standard uniform distribution, such that when multiplied by
sqrt(n) as part of the KS test it generates the desired metric. In R,
one can request asymptotic test via ks.test() with exact=F argument. Of
note, the /pKolmogorov() function in the cumulcalib package provides a
direct implementation of this test.

\begin{verbatim}

library(MASS)
data("birthwt")
pi <- 1/(1+exp(-(2.15-0.050*birthwt$age-0.015*birthwt$lwt))) #The model

#Ordering predictions (and responses) from smallest to largest pi
o <- order(pi)
pi <- pi[o]
Y <- birthwt$low[o] #Outcome is low birth weight (1 v 0)

#Constructing the {S} process
T <- sum(pi*(1-pi))
t <- cumsum(pi*(1-pi))/T
S <- cumsum(Y-pi)/sqrt(T)

plot(t,S, type='l')

Sn <- S[length(S)] #This is Sn, the first component of the bridge test
pA <- 2*pnorm(-abs(Sn)) #p-value for first component of H0

Bstar <- max(abs(S-t*Sn)) #This is B*, the second component of the bridge test
d <- Bstar/10
X <- seq(from=0, to=1-d, length.out=100)
pB <- ks.test(X, punif, exact=F)$p.value #p-value for second component of H0

#Fisher's method for generating a unified p-value
p <- 1-pchisq(-2*(log(pA)+log(pB)),4) #0.8381805

\end{verbatim}

\newpage

\end{document}